\documentclass[aps,prc,preprint,amsmath,amssymb,showpacs,preprintnumbers,superscriptaddress]{revtex4-1}
\usepackage{CJK}
\usepackage{graphicx}
\usepackage{dcolumn}
\usepackage{bm}
\usepackage{color}
\usepackage{amsmath}%
\allowdisplaybreaks[4]

\usepackage{hyperref}

\begin{document}

\title{Multiple chiral doublet bands with octupole correlations in reflection-asymmetric triaxial particle rotor model}

\author{Y. Y. Wang}
\affiliation{School of Physics and Nuclear Energy Engineering, Beihang University, Beijing 100191, China}
\author{S. Q. Zhang}\email{sqzhang@pku.edu.cn}
\affiliation{State Key Laboratory of Nuclear Physics and Technology, School of Physics, Peking University, Beijing 100871, China}
\author{P. W. Zhao}
\affiliation{State Key Laboratory of Nuclear Physics and Technology, School of Physics, Peking University, Beijing 100871, China}
\author{J. Meng}\email{mengj@pku.edu.cn}
\affiliation{State Key Laboratory of Nuclear Physics and Technology, School of Physics, Peking University, Beijing 100871, China}
\affiliation{School of Physics and Nuclear Energy Engineering, Beihang University, Beijing 100191, China}
\affiliation{Yukawa Institute for Theoretical Physics, Kyoto University, Kyoto 606-8502, Japan}

\date{\today}
\begin{abstract}
  A reflection-asymmetric triaxial particle rotor model (RAT-PRM) with a
  quasi-proton and a quasi-neutron coupled with a reflection-asymmetric
  triaxial rotor is developed and applied to investigate the multiple chiral
  doublet (M$\chi$D) bands candidates with octupole correlations in $^{78}$Br.
  The calculated excited energies, energy staggering parameters, and $B(M1)/B(E2)$
  ratios are in a reasonable agreement with the data of the chiral doublet bands
  with positive- and negative-parity. The influence of the triaxial deformation
  $\gamma$ on the calculated $B(E1)$ is found to be significant. By changing
  $\gamma$ from 16$^\circ$ to 21$^\circ$, the $B(E1)$ values will be enhanced and
  better agreement with the $B(E1)/B(E2)$ data is achieved. The chiral geometry
  based on the angular momenta for the rotor, the valence proton and valence neutron
  is discussed in details.
\end{abstract}
\maketitle

\date{today}

\section{Introduction}

Chirality is a subject of general interests in natural science. Since the
pioneering work of nuclear chirality by Frauendorf and Meng in 1997~\cite{Frauendorf1997Tilted},
many efforts have been devoted to explore the chirality in atomic nuclei, see e.g.,
reviews~\cite{Frauendorf2001Spontaneous,Meng2010Open,Meng2016Nuclear,Raduta2016Prog.Part.Nucl.Phys.241,
Starosta2017Phys.Scripta93002,Frauendorf2018Phys.Scripta43003}.

The experimental signature of nuclear chirality is a pair of nearly degenerate
$\Delta I=1$ bands with the same parity, i.e., chiral doublet bands. In 2006, the
multiple chiral doublets (M$\chi$D), i.e., more than one pair of chiral doublet
bands in a single nucleus, is suggested based on the self-consistent covariant
density functional theory (CDFT)~\cite{Meng2006Possible}. The first experimental
evidence for M$\chi$D is reported in $^{133}$Ce~\cite{Ayangeakaa2013Evidence},
followed by $^{103}$Rh \cite{Kuti2014Multiple}, $^{78}$Br~\cite{Liu2016Phys.Rev.Lett.112501},
$^{136}$Nd~\cite{Petrache2018Phys.Rev.C41304}, and $^{195}$Tl~\cite{Roy2018Phys.Lett.768}, etc.
Up to now, 62 candidate chiral bands in 49 nuclei (including 9 nuclei with M$\chi$D)
have been reported in the $A\sim$~80, 100, 130 and 190~mass regions
\cite{Xiong2019Atom.DataNucl.DataTabl.193,Wang2018Phys.Rev.14304,Roy2018Phys.Lett.768}.

Because of the observation of eight strong electric dipole ($E1$) transitions linking
the positive- and negative-parity chiral bands~\cite{Liu2016Phys.Rev.Lett.112501},
the M$\chi$D candidates observed in $^{78}$Br provide the first example of chiral
geometry in octupole soft nuclei and indicate that nuclear chirality can be robust
against the octupole correlations. It also indicates that the chirality-parity quartet bands~\cite{Frauendorf2001Spontaneous,Liu2016Phys.Rev.Lett.112501}, which are a consequence
of the simultaneous breaking of chiral and space-reflection symmetries, may exist in nuclei.
The observations of M$\chi$D with octupole correlations and/or the possible chirality-parity
quartet bands have brought severe challenges to current nuclear models and, thus,
require the development of new approaches.

Theoretically, nuclear chirality has been investigated extensively with
the triaxial particle rotor model (PRM)~\cite{Frauendorf1997Tilted,Peng2003Description,
Koike2004Chiral,Zhang2007Phys.Rev.C44307,Qi2009Chirality,Chen2018Phys.Lett.744},
the tilted axis cranking model (TAC)~\cite{Frauendorf1997Tilted,Dimitrov2000Chirality,
Olbratowski2004Critical,Olbratowski2006Search,Zhao2017Phys.Lett.B1}, the TAC approach
with the random phase approximation~\cite{Mukhopadhyay2007From,Almehed2011Chiral}
and the collective Hamiltonian~\cite{Chen2013Collective,Chen2016Two,Wu2018Phys.Rev.64302},
the interacting boson-fermion-fermion model~\cite{Brant2008Phys.Rev.34301}, the
generalized coherent state model \cite{Raduta2016J.Phys.95107} and the projected shell model \cite{Hara1995PROJECTED,Bhat2014Investigation,Chen2017Chiral,Chen2018Phys.Lett.211}, etc.
The triaxial PRM is one of the most popular models for describing nuclear chirality as it is
a quantal model coupling the collective rotation and the single-particle motions in the
laboratory reference frame, and describes directly the quantum tunneling and energy splitting
between the doublet bands.

In Ref.~\cite{Liu2016Phys.Rev.Lett.112501}, the triaxial PRM calculation has been performed
to describe the positive- and negative-parity chiral doublet bands observed in $^{78}$Br
with two individual configurations $\pi g_{9/2} \otimes \nu g_{9/2}$  and
$\pi f_{5/2} \otimes \nu g_{9/2}$, respectively. The calculation supports the interpretation
of the M$\chi$D with different parities~\cite{Liu2016Phys.Rev.Lett.112501}. However, the $E1$
linking transitions between the positive-parity band 1 and the negative-parity band 3 are not
accessible in the triaxial PRM due to the omission of the octupole degree of freedom.

In this work, a reflection-asymmetric triaxial PRM (RAT-PRM) with both triaxial and octupole
degrees of freedom is developed and applied to the M$\chi$D candidates with octupole
correlations in $^{78}$Br. The model is introduced in Sec.~\ref{sec1}, and the numerical
details are presented in Sec.~\ref{sec2}. The calculated results for the doublet bands,
such as energy spectra, electromagnetic transitions, and angular momentum orientations,
are discussed in Sec.~\ref{sec3}, and a summary is given in Sec.~\ref{sec4}.

\section{Formalism}\label{sec1}
The total RAT-PRM Hamiltonian can be  expressed as
\begin{align}
  \hat{H} = \hat{H}_{\rm intr.}^{p} + \hat{H}_{\rm intr.}^{n} + \hat{H}_{\rm core},
\end{align}
where $\hat{H}_{\rm intr.}^{p(n)}$ is the intrinsic Hamiltonian for valence protons (neutrons)
in a reflection-asymmetric triaxially deformed potential, and $\hat{H}_{\rm core}$ is the
Hamiltonian of a reflection-asymmetric triaxial rotor, which is generalized straightforwardly
from the reflection-asymmetric axial rotor in Ref.~\cite{Leander1984Nucl.Phys.375}.

The core Hamiltonian reads
\begin{align}
  \hat{H}_{\rm core}
= \sum_{k=1}^3\frac{\hat{R}_k^2}{2\mathcal{J}_k}
 +\frac{1}{2}E(0^-)(1-\hat{P}),
\end{align}
with $\hat{R}_k=\hat{I}_k-\hat{j}_{pk}-\hat{j}_{nk}$. Here, $\hat{R}_k$, $\hat{I}_k$, $\hat{j}_{pk}$,
and $\hat{j}_{nk}$ are the angular momentum operators for the core, the nucleus, the valence protons
and neutrons, respectively. The moments of inertia for irrotational flow are adopted $\mathcal{J}_k=\mathcal{J}_0\sin^2(\gamma-2k\pi/3)$. The core parity splitting parameter $E(0^-)$ can be viewed as the excitation energy of the virtual $0^-$ state~\cite{Leander1984Nucl.Phys.375}. The core parity operator $\hat{P}$ can be written as the product of the single-particle parity operator $\hat{\pi}$
and the total parity operator $\hat{p}$.

The intrinsic Hamiltonian for valence nucleons is~\cite{Hamamoto1976Nucl.Phys.15,Hamamoto1983Phys.Lett.281,
Zhang2007Phys.Rev.C44307}
\begin{align}
  \hat{H}_{\rm intr.}^{p(n)}
&=\hat{H}_{\rm s.p.}^{p(n)} + \hat{H}_{\rm pair}\notag\\
&=\sum_{\nu>0} (\varepsilon_\nu^{p(n)}-\lambda) (a_\nu^\dagger a_\nu+a_{\bar{\nu}}^\dagger a_{\bar{\nu}})
                 - \frac{\Delta}{2} \sum_{\nu>0}(a_{\nu}^\dagger a_{\bar{\nu}}^\dagger + a_{\bar{\nu}}a_\nu),
\end{align}
where $\lambda$ denotes the Fermi energy, $\Delta$ is the pairing gap parameter, and $|\bar{\nu}\rangle$ is
the time-reversal state of $|\nu\rangle$. The single-particle energy $\varepsilon_\nu^{p(n)}$ is obtained by
diagonalizing the Hamiltonian $\hat{H}_{\rm s.p.}^{p(n)}$, which has the form of a Nilsson Hamiltonian
\cite{nilsson1955see},
\begin{align}
  \hat{H}_{\rm s.p.}^{p(n)} = -\frac{1}{2}\hbar\omega_0\nabla^2 + V(r;\theta,\varphi)
                              + C\bm{l}\cdot \bm{s} +D[\bm{l}^2-\langle \bm{l}^2\rangle_N],
\end{align}
with the kinetic energy $-\frac{1}{2}\hbar\omega_0\nabla^2$, the reflection-asymmetric triaxially deformed
potential $V(r;\theta,\varphi)$ , the spin-orbit term $C\bm{l}\cdot \bm{s}$ and the standard
$D[\bm{l}^2-\langle\bm{l}^2\rangle_N]$ term~\cite{ring2004nuclear}.

Similar to Ref.~\cite{Hamamoto1991}, the reflection-asymmetric triaxially deformed potential
$V(r;\theta,\varphi)$ is written as
{\footnotesize
\begin{align}
V(r,\theta,\varphi)
=&\hbar\omega_0r^2
\Bigg[
  \frac{1}{2}+\beta_{10}Y_{10}
   + \beta_{11}\frac{(Y_{11}-Y_{1-1})}{\sqrt{2}}\notag\\
   &\quad\quad\quad\quad
   -\beta_{20} Y_{20}-\beta_{22}\frac{(Y_{22}+Y_{2-2})}{\sqrt{2}} \notag\\
   &\quad\quad\quad\quad
   -\beta_{30}Y_{30}-\beta_{31}\frac{(Y_{31}-Y_{3-1})}{\sqrt{2}}\notag\\
   &\quad\quad\quad\quad
   -\beta_{32}\frac{(Y_{32}+Y_{3-2})}{\sqrt{2}}-\beta_{33}\frac{(Y_{33}-Y_{3-3})}{\sqrt{2}}
\Bigg],
\end{align}}
with parameters $(\beta_{10},\beta_{11})$, $(\beta_{20},\beta_{22})$, and $(\beta_{30},\beta_{31},\beta_{32},\beta_{33})$ describing the dipole, quadrupole, and octupole deformations, respectively.
From the volume conservation and by requiring the center of mass coincided with the origin of the
coordinate system, the relations among the parameters can be obtained,
{\footnotesize
\begin{align}
 &\omega_0^2 \approx \Bigg[
          1+\frac{5}{16\pi}(\beta_{20}^2+\beta_{30}^2+\beta_{22}^2
           +\beta_{31}^2+\beta_{32}^2+\beta_{33}^2 )
          \Bigg]\mathring{\omega}_0^2,\\
 &\beta_{10} \approx \Bigg[ \frac{18\sqrt{3}}{\sqrt{35\pi}}\beta_{20}\beta_{30}
                        +\frac{6\sqrt{3}}{\sqrt{7\pi}}\beta_{22}\beta_{32}\Bigg],\\
 &\beta_{11}\approx \Bigg[
                        \frac{36}{\sqrt{70\pi}}\beta_{20}\beta_{31}
                       +\frac{18}{\sqrt{14\pi}}\beta_{22}\beta_{33}
                       - 12\sqrt{\frac{3}{280\pi}}\beta_{22}\beta_{31}
                   \Bigg].
\end{align}}
Here, the higher order terms of $(\beta_{10},\beta_{11})$, $(\beta_{20},\beta_{22})$, and $(\beta_{30},\beta_{31},\beta_{32},\beta_{33})$ are neglected, and $\mathring{\omega}_0$ corresponds to the frequency
of an equivalent spherical potential with conserved volume. The parameters $\beta_{20}$ and $\beta_{22}$
are related with the commonly used quadrupole deformation parameters $\beta_2$ and $\gamma$ by
\begin{align}
  \beta_{20}= \beta_2\cos\gamma,~~\beta_{22}=\beta_2\sin\gamma.
\end{align}

To include the pairing correlations in RAT-PRM, one should replace the single-particle states
$a_\nu^\dagger|0\rangle$ with the Bardeen-Cooper-Schrieffer (BCS) quasiparticle states
$\alpha_\nu^\dagger|\tilde{0}\rangle$, where $|\tilde{0}\rangle$ is the
BCS vacuum,
\begin{align}
  |\tilde{0}\rangle
= \prod_\nu (u_\nu+v_\nu a_\nu^\dagger a_{\bar{\nu}}^\dagger)|0\rangle,
\end{align}
and the quasiparticle operators $\alpha_\nu^\dagger$ read
\begin{align}
  \begin{pmatrix}
    \alpha_\nu^\dagger \\
    \alpha_{\bar{\nu}}
  \end{pmatrix}
  =
  \begin{pmatrix}
    u_\nu & -v_\nu\\
    v_\nu & u_\nu
  \end{pmatrix}
  \begin{pmatrix}
    a_\nu^\dagger \\
    a_{\bar{\nu}}
  \end{pmatrix},
\end{align}
with $u_\nu^2+v_\nu^2=1$. Furthermore, the single-particle energies $\varepsilon_\nu$ should be
replaced by quasiparticle energies $\varepsilon_\nu'=\sqrt{(\varepsilon_\nu-\lambda)^2+\Delta^2}$.
Therefore, the intrinsic Hamiltonian becomes
\begin{align}
  \hat{H}_{\rm intr.}
=\sum_{\nu_p} \varepsilon'_{\nu_p} ( \alpha_{\nu_p}^\dagger \alpha_{\nu_p}
                                       +\alpha_{\bar{\nu}_p}^\dagger \alpha_{\bar{\nu}_p})
 +\sum_{\nu_n} \varepsilon'_{\nu_n} ( \alpha_{\nu_n}^\dagger\alpha_{\nu_n}
                                        +\alpha_{\bar{\nu}_n}^\dagger \alpha_{\bar{\nu}_n}).
\end{align}

The Hamiltonian $\hat{H}$ is diagonalized numerically in the symmetrized strong-coupled basis with
good parity and angular momentum,
\begin{align}
  |\Psi_{IMK\pm}^\nu\rangle
= \frac{1}{2\sqrt{1+\delta_{K0}}} (1+\hat{S}_2)|IMK\rangle \psi_{\pm}^\nu,
\end{align}
where $\hat{S}_2=\hat{P}\hat{R}_2$ is the reflection operator with respect to the plane perpendicular
to 2-axis, $|IMK\rangle = \sqrt{\displaystyle\frac{2\pi+1}{8\pi^2}}D_{MK}^{I*}$ is the Wigner function,
$\psi_{\pm}^\nu$ are the intrinsic wavefunctions with good parity,
\begin{align}
  \psi_+^\nu & =  (1+\hat{p})\tilde{\chi}_p^\nu\tilde{\chi}_n^\nu\Phi_a
               =  (1+\hat{P}\hat{\pi}_p\hat{\pi}_n)\tilde{\chi}_p^\nu\tilde{\chi}_n^\nu\Phi_a,\\
  \psi_-^\nu & =  (1-\hat{p})\tilde{\chi}_p^\nu\tilde{\chi}_n^\nu\Phi_a
               =  (1-\hat{P}\hat{\pi}_p\hat{\pi}_n)\tilde{\chi}_p^\nu\tilde{\chi}_n^\nu\Phi_a.
\end{align}
Here $\tilde{\chi}_p^\nu\tilde{\chi}_n^\nu\Phi_a$ is the strong-coupled intrinsic core-quasiparticle
wavefunction; $\Phi_a$ represents that the core has the same orientation in space as the intrinsic
single-particle potential, and $\tilde{\chi}_{p(n)}^\nu$ is the BCS quasiparticle state of the proton
(neutron).

The diagonalization of the Hamiltonian $\hat{H}$ gives rise to the nuclear eigenstate,
\begin{align}
  |IMp\rangle = \sum_{\nu K} c_{IKp}^\nu |\Psi_{IMKp}^\nu\rangle,\quad p=\pm,
\end{align}
which is a composition of the strong-coupled basis with the coefficients $c_{IKp}^\nu$. Then, the reduced
electromagnetic transition probabilities can be calculated via~\cite{bohr1975nuclear}
\begin{align}
   B(\sigma\lambda, I_i\rightarrow I'_f)
=\frac{1}{2I+1}\sum_{\mu M'}|\langle I'M'p'|\mathcal{M}_{\lambda\mu}^\sigma|IMp\rangle|^2,
\end{align}
where $\sigma$ denotes either $E$ or $M$ for electric and magnetic transitions, respectively, $\lambda$
is the rank of the transition operator, and $\mathcal{M}_{\lambda\mu}^\sigma$ the electromagnetic
transition operator.

The magnetic dipole ($M1$) transition operator is
\begin{align}
  \hat{\mathcal{M}}(M1,\mu)
 =\sqrt{\frac{3}{4\pi}}\frac{e\hbar}{2Mc}\Bigg[(g_p-g_R)\hat{j}^p_{1\mu}+(g_n-g_R)\hat{j}^n_{1\mu}\Bigg],
\end{align}
where $g_p$, $g_n$, and $g_R$ are the effective gyromagnetic ratios for valence proton, valence neutron,
and the collective core, respectively, and $\hat{j}_{1\mu}$ denotes the spherical tensor in the laboratory
frame. The electric multipole transition operators contain two terms~\cite{ring2004nuclear},
\begin{align}
     \hat{\mathcal{M}}(E\lambda,\mu)
    &= \hat{q}_{\lambda\mu}^{(c)} + \hat{q}_{\lambda\mu}^{(p)}\notag\\
    &= \frac{3Ze}{4\pi}R_0^\lambda \beta_{\lambda\mu}
      + e\sum_{i=1}^n (\frac{1}{2}-t_3^{(i)})r_i^\lambda Y_{\lambda\mu}^*,
\end{align}
which are contributions from the core and the valence particles, respectively. Here, $R_0 = 1.2 A^{1/3}$
is the nuclear radius. For electric quadrupole ($E2$) transitions, one can safely neglect the valence
particle term, since it is much smaller than the term of the core~\cite{ring2004nuclear}. However, this
is not the case for $E1$ transitions. Since the total center of mass remains at rest, the motion of the
valence particles is influenced by the recoil of the core. This effect is of special importance for
$E1$ transitions. Therefore, as in Ref.~\cite{bohr1975nuclear}, the total moment in a one-particle
transition is obtained by replacing the charge of the particle by an effective one,
\begin{align}
  e_{\rm eff}
  = (\frac{1}{2}-t_3^{(i)})e-\frac{Ze}{A}
  =\begin{cases}
    \frac{N}{A}e \quad {\rm ~~for~proton,~}t_3^{i}=-\frac{1}{2}\\
    -\frac{Z}{A}e \quad {\rm for~neutron,~}t_3^{i}=\frac{1}{2}
  \end{cases}.
\end{align}

\section{Numerical details}\label{sec2}
The microscopic multidimensionally-constrained covariant density functional theory (MDC-CDFT)~\cite{Lu2012Phys.Rev.11301,Zhao2012Phys.Rev.57304,Lu2014Phys.Rev.15} with
PC-PK1~\cite{Zhao2010Phys.Rev.C54319} gives the quadrupole deformation $\beta_2=0.28$,
$\gamma=16.3^\circ$, and the octupole deformation $\beta_3=0$ for the configuration
$\pi g_{9/2}\otimes \nu g_{9/2}$ in $^{78}$Br. As the potential energy surface is soft
with respect to $\beta_3$, $\beta_3=0.02$ is adopted to include the effect of octupole
correlations in the present RAT-PRM calculations.

With the deformation parameters above, the reflection-asymmetric triaxial Nilsson Hamiltonian
with the parameters $\kappa,\mu$ in Ref.~\cite{nilsson1969nuclear} is solved by expanding the
wavefunction by harmonic oscillator basis~\cite{Wang2018Sci.ChinaPhys.Mech.Astron.82012}. The
Fermi energies of proton and neutron are chosen as $\lambda_p=44.6$ MeV and $\lambda_n=47.6$ MeV,
corresponding to the $\pi g_{9/2}[m_z=1/2]$ and $\nu g_{9/2}[m_z=5/2]$ orbitals respectively,
which are consistent with the MDC-CDFT results. The single-particle space is truncated to 13
levels, with six above and below the Fermi level. Increasing the size of the single-particle
space does not influence the band structure in the present work. The pairing correlation is
taken into account by the empirical pairing gap formula $\Delta = 12/\sqrt{A}$ MeV.

The moment of inertia $\mathcal{J}_0=14$~$\hbar^2/$MeV and the core parity splitting parameter
$E(0^-)=3$~MeV, are adjusted to the experimental energy spectra. For the calculations of magnetic
transitions, the gyromagnetic ratios for the collective rotor, protons, and neutrons are given by
$g_R=Z/A$, $g_{p(n)}=g_l+(g_s-g_l)/(2l+1)$, respectively~\cite{bohr1975nuclear,ring2004nuclear}.

\section{Results and discussion}\label{sec3}
\begin{figure}[!htbp]
  \centerline{
  \includegraphics[width=0.4\textwidth]{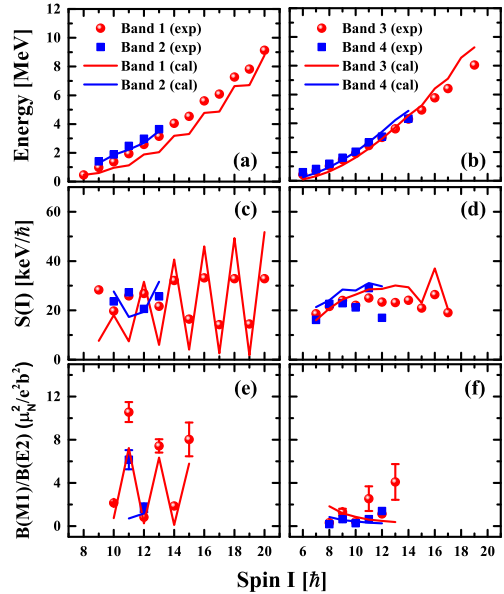}}
  \caption{ The excitation energies [panels (a) and (b)], the energy staggering parameters
  $S(I)=[E(I)-E(I-1)]/2I$ [panels (c) and (d)], and the $B(M1)/B(E2)$ ratios [panels (e) and (f)]
  for the positive-parity doublet bands 1 and 2 (left panels) as well as the negative-parity
  doublet bands 3 and 4 (right panels) in $^{78}$Br by RAT-PRM (lines) in comparison with the
  data available~\cite{Liu2016Phys.Rev.Lett.112501} (symbols). The energy of band 1 at $I=8\hbar$
  is renormalized to the corresponding experimental bandhead.}
  \label{fig1}
\end{figure}
In Fig.~\ref{fig1}, the excitation energies, the energy staggering parameters $S(I)=[E(I)-E(I-1)]/2I$,
and the $B(M1)/B(E2)$ ratios calculated by the RAT-PRM for the positive-parity doublet bands 1 and 2
as well as the negative-parity doublet bands 3 and 4 are shown in comparison with the data available
\cite{Liu2016Phys.Rev.Lett.112501}.

As shown in Figs.~\ref{fig1}(a) and~\ref{fig1}(b), the calculated excited energies reproduce the data
for the positive-parity doublet bands satisfactorily, and for the negative-parity doublet bands very
well. Within the spin region $9\hbar \le I \le 13\hbar$, the average energy difference for the
positive-parity doublet bands is 0.99 MeV, which overestimates the data by $\sim$0.5 MeV. Within the
spin region $6\hbar \le I \le 14\hbar$, the average energy difference for the negative-positive doublet
bands is 0.35 MeV, which overestimates the data by $\sim$0.2 MeV. The overestimation of the energy
splittings between doublet bands may be due to the small triaxial deformation ($\gamma=16.3^\circ$) adopted
in the present calculations. In Ref.~\cite{Landulfo1996Phys.Rev.626}, the cranked-shell-model calculations
suggest the deformation parameters $(\beta_2,\gamma)=(0.32,21.3^\circ)$ for band 1 in order to match the
experimental moments of inertia. The tilted axis cranking CDFT (TAC-CDFT) calculations~\cite{Zhao2011Phys.Lett.B181,Zhao2011Phys.Rev.Lett.122501,Meng2013Front.Phys.55,
Zhao2015Phys.Rev.Lett.22501,Zhao2017Phys.Lett.B1,Zhao2018IJMPE} indicate that the triaxial deformation
increases with the rotational frequency. By using a larger triaxial deformation, the RAT-PRM calculations
could provide smaller average energy difference for the positive-parity doublet bands and the negative-parity
doublet bands.

Figs.~\ref{fig1}(c) and~\ref{fig1}(d) depict the calculated $S(I)$ values in comparison with the data. For
the positive-parity doublet bands, the calculated $S(I)$ values exhibit an odd-even staggering behavior.
For the negative-parity doublet bands, the calculated $S(I)$ values are smooth till $14\hbar$. The different
$S(I)$ behaviors may be attributed to their corresponding configurations. The proton configurations are
similar for both positive- and negative-parity bands, i.e., a particle at the bottom of the $g_{9/2}$ shell.
The neutron configurations, however, are quite different. There is a neutron hole at the top of the $f_{5/2}$
shell for the negative-parity bands, but a neutron at the middle of the $g_{9/2}$ shell for the positive-parity
bands. For the latter, the neutron alignments along the direction of the collective rotation may occur, and
the $S(I)$ staggering appears.

The experimental $B(M1)/B(E2)$ ratios for both positive- and negative-parity doublet bands, including the odd-even
staggering for the positive-parity, are well reproduced, as shown in Figs.~\ref{fig1}(e) and~\ref{fig1}(f).
The similarity of $B(M1)/B(E2)$ ratios between the doublet bands is an indication for nuclear chirality as suggested in Ref.~\cite{Wang2007Examining}.

\begin{figure}[!htbp]
  \centerline{
  \includegraphics[width=0.42\textwidth]{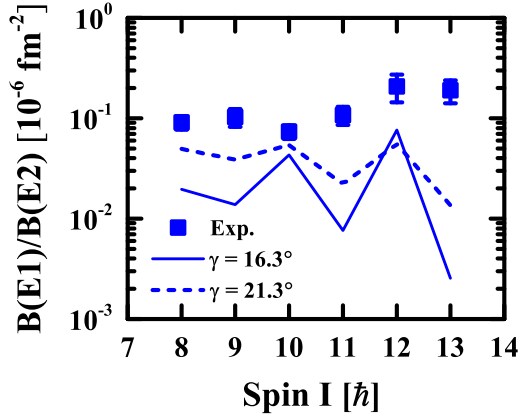}}
  \caption{The calculated $B(E1)/B(E2)$ ratios, with the interband $E1$ transitions (band 3 $\rightarrow$ 1) and
   the intraband $E2$ transitions (band 3), in comparison with the available data~\cite{Liu2016Phys.Rev.Lett.112501}.
   The solid and dashed lines represent the results calculated by RAT-PRM with triaxial deformations
   $\gamma=16.3^\circ$ and $21.3^\circ$, respectively. }
  \label{fig2}
\end{figure}

Since the octupole degree of freedom is included in the present RAT-PRM calculations, the electric
dipole transition probabilities $B(E1)$ between the positive- and negative-parity bands can be calculated. In Fig.
\ref{fig2}, the calculated $B(E1)/B(E2)$ ratios with the interband $E1$ transitions (band 3 $\rightarrow$ 1) and
the intraband $E2$ transitions (band 3) are shown in comparison with the available data~\cite{Liu2016Phys.Rev.Lett.112501}.
In general, the calculated $B(E1)/B(E2)$ ratios underestimate the experimental data. Considering the fact that the
calculated $B(M1)/B(E2)$ ratios for band 3 agree with the data, the underestimation of the calculated $B(E1)/B(E2)$
ratios may result from too small $B(E1)$ values.

It is found that the influence of the triaxial deformation $\gamma$ on the calculated $B(E1)$ is significant.
By changing $\gamma$ from 16$^\circ$ to 21$^\circ$ (given by cranked-shell-model calculations~\cite{Landulfo1996Phys.Rev.626}),
as shown in Fig.~\ref{fig2}, the $B(E1)$ values will be enhanced and better agreement with the $B(E1)/B(E2)$ data is achieved.

\begin{figure}[!htbp]
  \centerline{
  \includegraphics[width=0.5\textwidth]{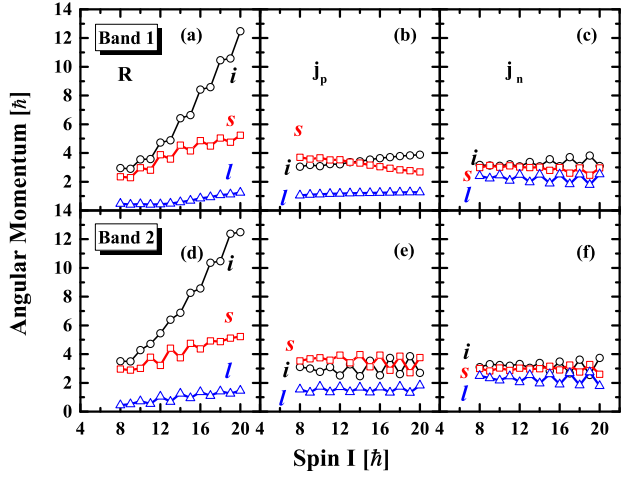}}
  \caption{The angular momenta components along the intermediate ($i$, circles), short ($s$, squares),
  and long ($l$, triangles) axes for the core $R_k=\langle \hat{R}_k^2\rangle^{1/2}$ [panels (a) and (d)],
  valence proton $j_{pk}=\langle \hat{j}_{pk}^2\rangle^{1/2}$ [panels (b) and (e)], and valence neutron
  $j_{nk}=\langle \hat{j}_{nk}^2\rangle^{1/2}$ [panels (c) and (f)] in RAT-PRM for the positive-parity
  doublet bands 1 and 2.}
  \label{fig3}
\end{figure}
\begin{figure}[!htbp]
  \centerline{
  \includegraphics[width=0.43\textwidth]{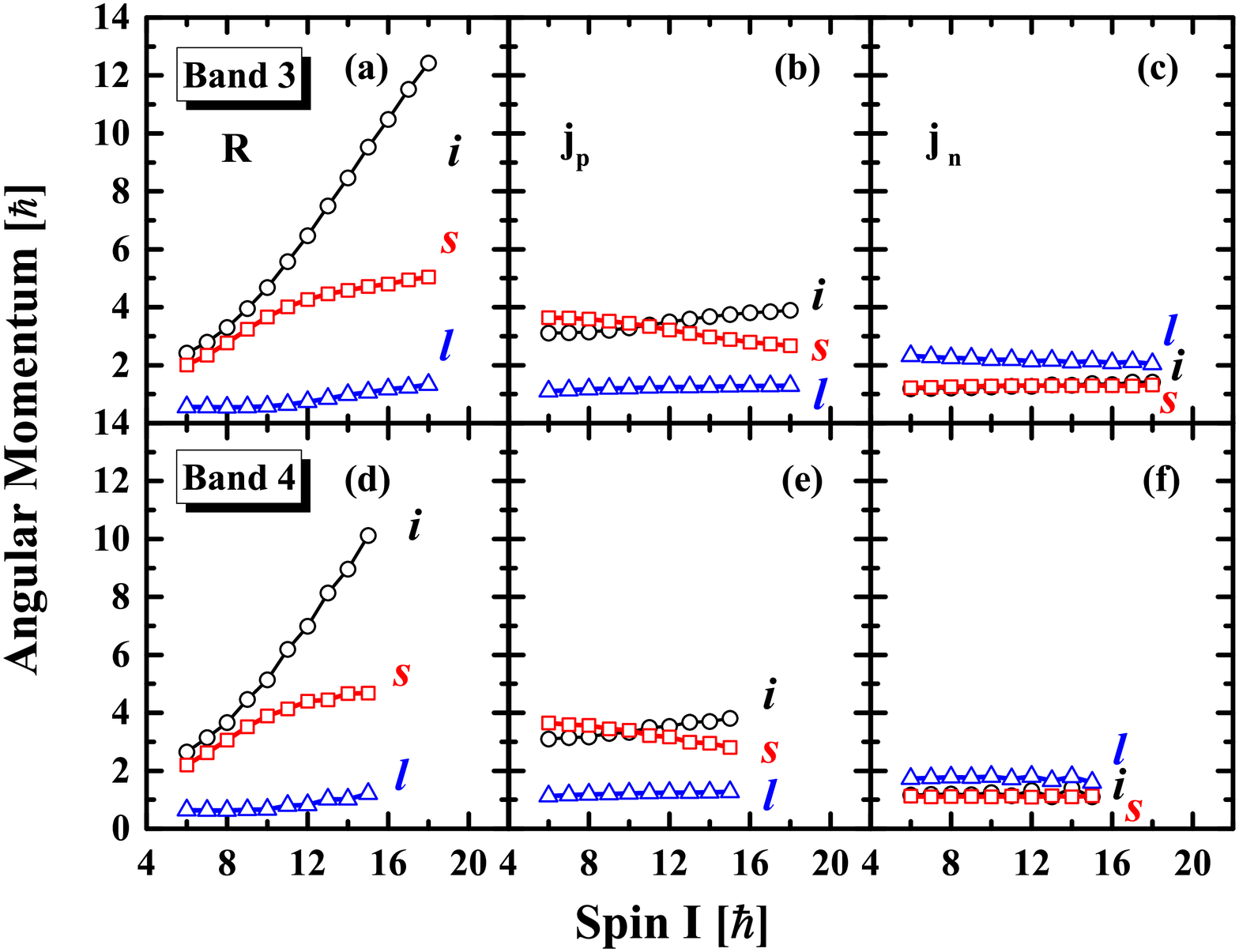}}
  \caption{ Same as Fig.~\ref{fig3}, but for the negative-parity doublet bands 3 and 4. }
  \label{fig4}
\end{figure}

In order to investigate the chiral geometry, the angular momentum components for the core $R_k=\langle \hat{R}_k^2\rangle^{1/2}$,
the valence proton $j_{pk}=\langle \hat{j}_{pk}^2\rangle^{1/2}$, and the valence neutron $j_{nk}=\langle \hat{j}_{nk}^2\rangle^{1/2}
(k=1,2,3)$ are presented in Figs.~\ref{fig3} and \ref{fig4} for the positive- and negative-parity doublet bands, respectively.
For the triaxial deformation $\gamma=16.3^\circ$ adopted here, the intrinsic axes 1, 2, and 3 are respectively the intermediate
($i$), short ($s$) and long ($l$) axes, and the relation of the corresponding moments of inertia is $\mathcal{J}_1>\mathcal{J}_2>
\mathcal{J}_3$. Therefore, as shown in Figs.~\ref{fig3} and \ref{fig4}, the angular momentum for the core mainly aligns along the
$i$-axis for both positive- and negative-parity doublet bands.

For the positive-parity doublet bands in Fig.~\ref{fig3}, the angular momentum of the valence proton mainly aligns in the $i$-$s$ plane,
while that of the valence neutron has nearly equal components on the three axes due to its mid-shell nature. Considering the fact that
the angular momentum for the core mainly aligns along the $i$-axis, and grows rapidly, the total angular momentum lies close to the
$i$-$s$ plane, which is consistent with the large energy difference between the doublet bands. For band 1, the three components of $j_p$
for the valence proton vary smoothly with the spin, while the three components of $j_n$ for the valence neutron exhibit staggering with
$I>12\hbar$. For band 2, the three components of both $j_p$ and $j_n$ exhibit staggering with $I\geq 9\hbar$. These staggering behaviors
might be understood from the main components of the intrinsic wavefunction $\tilde{\chi}_{p(n)}$. It is found that these staggering
behaviors are associated with the variation of the corresponding main components. Taking band 1 as an example, with $I>12\hbar$, the main
component of the neutron intrinsic wavefunction varies alternately between $g_{9/2}[m_z=5/2]$ and $g_{9/2}[m_z=3/2]$.

For the negative-parity doublet bands in Fig.~\ref{fig4}, the angular momentum of the valence proton mainly aligns in the $i$-$s$ plane,
and the alignment of the valence neutron along the $l$-axis is significant. To be more precise, $j_p\sim 4\hbar$ in the $i$-$s$ plane,
$j_n\sim 2\hbar$ along $l$-axis, and $R\sim 2\hbar-13\hbar$ along $i$-axis. This is the chiral geometry for the negative-parity doublet
bands. As the total angular momentum increases, $R$ increases gradually, $j_n$ remains almost unchanged, while $j_p$ moves gradually toward the
$i$-axis. The difference between the proton and neutron alignments may result from the fact that the Coriolis alignment effects are weaker
for the neutron in the relatively low-$j$ $f_{5/2}$ shell. It is found that for both band 3 and band 4, the three components of $j_p$ and
$j_n$ vary smoothly with the spin. This is different from the case of band 1 and band 2, because the main components here are always
$g_{9/2}[m_z=1/2]$ for proton and $f_{5/2}[m_z=5/2]$ for neutron.

\begin{figure}[!htbp]
  \centerline{
  \includegraphics[width=0.43\textwidth]{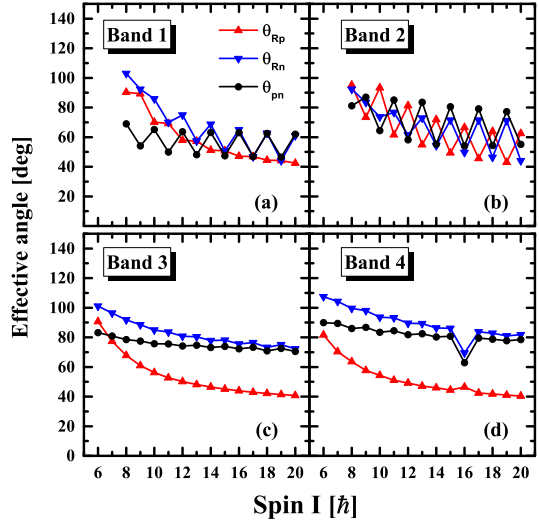}}
  \caption{The effective angles $\theta_{Rp}$ (triangle ups), $\theta_{Rn}$ (triangle downs) and $\theta_{pn}$ (circles) for the
  positive- [panels (a) and (b)] and negative-parity doublet bands [panels (c) and (d)] as functions of spin.}
  \label{fig5}
\end{figure}
In Fig.~\ref{fig5}, the calculated effective angles $\theta_{Rp}$, $\theta_{Rn}$ and $\theta_{pn}$  as functions of spin for the
positive- and negative-parity doublet bands are presented. The effective angle $\theta_{pn}$ between the angular momenta of the
proton $\bm{j}_p$ and neutron $\bm{j}_n$ is defined as~\cite{Starosta2002Phys.Rev.44328}
 \begin{align}
   \cos \theta_{pn}
 =\frac{\langle \bm{j}_p\cdot\bm{j}_n\rangle}{\sqrt{\langle j_p^2\rangle\langle j_n^2\rangle}}.
 \end{align}
A similar expression for the effective angle $\theta_{Rp}(\theta_{Rn})$ between the angular momenta for the core and the valence
proton (neutron) can be defined straightforwardly.

In Fig.~\ref{fig5}, the three effective angles for both the positive- and negative-parity doublet bands decrease with spin. This
behavior can be well understood because both the valence proton and the valence neutron gradually align along the direction of
the collective rotation with spin. For the positive-parity doublet bands, $\theta_{Rp}$ and $\theta_{Rn}$ decrease with spin with
almost the same slope. This is because the Coriolis alignment effects for the valence proton in $\pi g_{9/2}$ and the valence
neutron in $\nu g_{9/2}$ are similar. In contrast, for the negative-parity doublet bands, $\theta_{Rp}$ decreases faster than
$\theta_{Rn}$ with spin. This is because the Coriolis alignment effects for the valence proton in $g_{9/2}$ are stronger than
those for the valence neutron in relatively low-$j$ $f_{5/2}$ orbitals.

The effective angles oscillate with spin for the positive-parity doublet bands, while smoothly change with spin for the negative-parity
doublet bands except a kink at $I=16\hbar$ in band 4. The staggering features for bands 1 and 2 are again connected with the change
for the main component in the intrinsic wavefunctions, as discussed above. For bands 3 and 4, the main components in the intrinsic
wavefunctions, $\pi g_{9/2}[m_z=1/2]$ and $\nu f_{5/2}[m_z=5/2]$, are nearly unchanged, which result in the smooth change of the
effective angles. In addition, the kink at $I=16\hbar$ for band 4 is due to a sudden change of the main component in the intrinsic
wavefunctions, from $\pi g_{9/2}\otimes \nu f_{5/2}$ to $\pi g_{9/2}\otimes \nu g_{9/2}$.

Finally, a few remarks on the effective angles near the bandheads are appropriate. In Ref.~\cite{Chen2018Phys.Rev.31303}, the paradox,
i.e., the effective angles between any two of the angular momentum components are closed to $90^\circ$ in the regime of chiral vibration,
has been clarified. This paradox is due to the fact that the angular momentum of the rotor is much smaller than those of the proton and
neutron near the bandhead. Here, the three effective angles near the bandheads, in the regime of chiral vibration,
are close to $90^\circ$ for positive-parity band 2, and negative-parity doublet bands 3 and 4. However, the effective angle $\theta_{pn}$
at the bandhead for band 1 is only $70^\circ$ due to the deviation from the ideal particle-hole configuration and triaxial deformation.

It should be noted that for the negative-parity doublet bands the previous adopted configuration is $\pi f_{5/2}\otimes \nu g_{9/2}$~\cite{Liu2016Phys.Rev.Lett.112501}.
In the present calculations, the configuration $\pi g_{9/2}\otimes \nu g_{9/2}$
same as in Ref.\cite{Liu2016Phys.Rev.Lett.112501} is adopted for the positive-parity doublet bands. After including the
octupole deformation, the positive- and negative-parity bands can be simultaneously obtained by diagonalizing the RAT-PRM Hamiltonian.
For the yrast band with negative parity, the configuration is found to be $\pi g_{9/2}\otimes \nu f_{5/2}$. Further support for this subtle change in configuration for the negative-parity doublet bands may be obtained from future microscopic calculations and experimental results, for example, the three-dimensional TAC-CDFT~\cite{Zhao2017Phys.Lett.B1} including the octupole deformation or the measurement of the $g$ factor in the chiral bands~\cite{Grodner2018Phys.Rev.Lett.22502a}.

\section{Summary}\label{sec4}{}

In summary, a reflection-asymmetric triaxial particle rotor model (RAT-PRM) with a quasi-proton and a quasi-neutron coupled with a reflective-asymmetric triaxial rotor is developed and applied to the M$\chi$D candidates with octupole correlations in $^{78}$Br.

The excited energies, energy staggering parameters $S(I)=[E(I)-E(I-1)]/2I$ and $B(M1)/B(E2)$ ratios are calculated for the
positive-parity doublet bands 1 and 2 as well as the negative-parity doublet bands 3 and 4. Since the octupole deformation
is included in the present RAT-PRM calculations, the electric dipole transition probabilities $B(E1)$ between the positive-
and negative-parity bands can be calculated. The calculated excited energies and the energy staggering parameters reproduce
the data for the positive-parity doublet bands satisfactorily, and for the negative-parity doublet bands very well. The calculated
$B(M1)/B(E2)$ ratios agree well with the experimental data, while the calculated $B(E1)/B(E2)$ ratios underestimate the data in
general. It is found that the influence of the triaxial deformation $\gamma$ on the calculated $B(E1)$ is significant.
By changing $\gamma$ from 16$^\circ$ to 21$^\circ$ (given by cranked-shell-model calculations~\cite{Landulfo1996Phys.Rev.626}),
the $B(E1)$ values will be enhanced and better agreement with the $B(E1)/B(E2)$ data is achieved.

The chiral geometry and its evolution are discussed in details from the angular momentum components for the core as well as the
valence proton and neutron. For the positive-parity doublet bands, in consistent with the large energy difference between the
doublet bands, the total angular momentum lying close to the $i$-$s$ plane. For the negative-parity doublet bands, the chiral
geometry is constructed by the angular momenta of the valence proton along the $i$-$s$ plane, the valence neutron along the
$l$-axis, and the core along the $i$-axis.

\begin{acknowledgments}
This work was partly supported by the National Key R\&D Program of China
(Contract No. 2018YFA0404400 and No. 2017YFE0116700), the National Natural
Science Foundation of China (Grants No. 11621131001 and No. 11875075).
\end{acknowledgments}

\end{document}